\documentclass[11pt]{amsart}
\newtheorem{theorem}{Theorem}[section]

\theoremstyle{definition}

\theoremstyle{remark}

\numberwithin{equation}{section}

\newcommand{\HH}{{\mathcal{H}}}

\begin{document}

\title[Gelfand-Shilov spaces]
{Gelfand-Shilov spaces\\ Structural and Kernel theorems }

\author{Z. Lozanov--Crvenkovi\'c}
\address{Department  of Mathematics and Informatics,
University of Novi Sad, Trg Dositeja Obradovi\'{c}a 4, 21000 Novi
Sad, Serbia}\email{zlc@im.ns.ac.yu}

\author{D.Peri\v{s}i\'{c} }
\address{Department  of Mathematics and Informatics,
University of Novi Sad, Trg Dositeja Obradovi\'{c}a 4, 21000 Novi
Sad, Serbia} \email{dusanka@im.ns.ac.yu}

\author{M.Taskovi\'{c}}
\address{Department  of Mathematics and Informatics,
University of Novi Sad, Trg Dositeja Obradovi\'{c}a 4, 21000 Novi
Sad, Serbia} \email{m.tasko@EUnet.yu}





\begin{abstract}In the paper we give the structural (regularity) theorem and kernel theorem
for  Gelfand-Shilov spaces,  of Roumieu and Beurling type.

\end{abstract}

\maketitle

\section{Introduction}

The  Gelfand-Shilov spaces of Roumieu and Beurling type, denoted
by ${{\mathcal S}^{\{M_p\}}}$ and ${{\mathcal S}^{(M_p)}}$, are
subclasses of Denjoy-Carleman classes $C^{\{M_p\}}$  and
$C^{(M_p)}$, which are invariant under Fourier transform,  closed
under differentiation and multiplication by polynomials, and which
contain Schwartz space of tempered distributions as a subspace.
This makes the Gelfand Shilov spaces appropriate domains for
harmonic analysis. The space ${{\mathcal S}^{\{M_p\}}}$ (resp.\
${{\mathcal S}^{(M_p)}}$) is defined as the inductive (resp.\
projective) limit of Banach spaces ${\mathcal S}^{M_p,m}, $ $m
>0$, where by ${\mathcal S}^{M_p,m} $,  we denote the  space of
smooth functions $\varphi $ on ${\Bbb R}^{d}$, such that for some
$C>0$ and $m>0$
\begin{equation}\label{M11}
||\varphi||_{{\mathcal S}^{M_p},m} = \sup_{\alpha,\beta \in {\Bbb
N}_{0}^{d}} \frac{m^{|\alpha|+|\beta|}}{M_{|\alpha|}M_{|\beta|}}
||(1+x^2)^{\beta/2}\varphi^{(\alpha)}(x)||_{L^{\infty}}< \infty,
\end{equation}
and is equipped with the norm $|| \cdot||_{{\mathcal S}^{M_p},m}.
$ \\

Examples of classes of sequences, which generate the spaces and
 satisfy the conditions, we assume in the paper, are:
$$
M_p=e^{p^r},\;\;\; p \in {\Bbb N},\quad r\in(1,2].
$$
and
\begin{equation}\label{Mp}
M_p = p^{sp}(\log  p )^{tp}, \; p \in {\Bbb N},\; s\geq 1/2,\;t
\geq 0, \\
\end{equation}
wtih additional condition (only) in Beurling case: $s+t>1/2$.

 So
the classes of sequences $\{M_p\}$, which we consider, are much
wider that just
the class of Gevery sequences. \\

Many classical spaces of analysis are Gelfand Shilov spaces. In
the special case when $\{M_p\}$ is a Gevrey sequence
$p!^{\alpha}$,  the space is equal to ${\mathcal
S}^{\alpha}_{\alpha}$, $\alpha\geq 1/2$ (\cite{Gelfand2}). The
dual of this space has been successfully used in differential
operators theory, spectral analysis, and more recently in theory
of pseudodifferential operators (\cite{TeofanovJFA}) and in
quantum field theory. It was shown (\cite{Fizika},\cite{Fizika2})
that the space isomorphic with ${\mathcal S}^{\{p!\}}={\mathcal
S}^1_1$ is well adapted for the use in quantum field theory with a
fundamental length . It is our believe that all Gelfand Shilov
spaces, especially those with quasianalytic test function spaces,
are good domains for  the quantum field theory. The theory
requires technical results from the theory of generalized
functions and not merely differential calculus and well defined
Fourier transform, but also the kernel theorem and
the structural  theorem.\\

The aim of this paper is to present simple proofs of the kernel
and structural theorem for the Gelfand Shilov spaces (in
quasianalytic and nonquasianalytic case in an uniform way) using
the minimal amount of real analysis. Recently published results
giving characterizations of Fourier Hermite coefficients of
Gelfand Shilov spaces, enabled us to use  Simons ideas
(\cite{Simon}), and use the harmonic oscillator wavefunctions, as
a basic tool for proving kernel and structural theorem.  The
characterization has been given by Langenbruch in \cite{L}.
Without knowing for the result, and unfortunately after the
Langenbruch paper was published online, we (Perisic and
Lozanov-Crvenkovic) have obtained (\cite{nas rad})
the same characterization, but using different techniques.  \\

  In Section \ref{main
results}  we prove,  that every element $T$ of
 the dual space ${{\mathcal S}^{\{M_p\}}}'$ (resp.\ ${{\mathcal
S}^{(M_p)}}'$ ) is of the form:
$$
T = \sum_{\beta = 0}^{\infty} \frac{\mu^{2\beta}}{M_{2\beta}}
    \frac{1}{2^{\beta}} \left(- \frac{d^2}{dx^2}+x^2 -1
    \right)^{\beta}f
$$
for every  (resp.\ some) $\mu > 0$ and some continuous bounded
function $f$, (i.e. structural theorem ).\\

In the same Section we prove the kernel theorem, which states
every continuous linear map ${\mathcal K}$ on the space
$({\mathcal S}^{\{M_p\}}({\Bbb R}^l))_x$ of test functions in some
variable $x$, into the space $({{\mathcal S}^{\{M_p\}}}'({\Bbb
R}^s))_y$ of in a second variable $y$, is given by a unique $K \in
({{\mathcal S}^{\{M_p\}}}'({\Bbb R}^{l+s}))_{x,y}$ in both
variables $x$ and $y$. \\

An immediate consequence of the Kernel theorem is that the
representation of the Heisenberg group and the Weyl transform can
be extended to the Gelfand-Shilov spaces generated by a sequences
$\{M_p\}$ from a large class of sequences, where the Gevrey
sequence (either in quasianalitic or nonquasianlitic case) is just
one of the examples. This
 gives the possibility to introduce the $\Psi$DO theory
 in the framework of the spaces.
In \cite{TeofanovJFA} it was demonstrated how it can be done by
using Gelfand-Shilov spaces of Beurling type generated by Gevrey
sequences $\{p!^{\alpha}\}_p$, for $\alpha\geq 1$, that is in
strong
nonquasianalityc case.\\

In Section \ref{remarks} we   list several examples of
Gelfand-Shilov spaces,  discus conditions we assume on the
sequences $\{ M_p\}$ and their consequences on the structure of
the Gelfand-Shilov spaces.

\subsection{Notations and basic notions}

Throughout  the paper by $C$ we denote a positive constant, not
necessarily the same at each occurrence. Also we use $*$ instead
of
$\{M_p\}$ or $(M_p)$.\\

Let $ \{M_{p}\}_{p \in {\Bbb N}_{0}}$ be a sequence of positive
numbers, where $M_{0} =1$. We also assume that the sequence
$\{M_{p}\}_{p \in  {\Bbb N}_{0}}$ satisfies:
\begin{itemize}
\item[(M.1)]
$ M^{2}_{p} \leq M_{p-1} M_{p+1}, \quad  p = 1,2,\dots . $\\ ({\it
logarithmic convexity})
\\
\item[(M.2)]
{\it There exist constants $A$, $H>0$ such that}\\
 $ M_{p} \leq  A H^{p}
\min_{0 \leq q \leq p}M_{q}M_{p-q} ,
     \quad p = 0,1,\dots  \;
$\\
({\it separativity condition  or stability under ultradifferential
operators})
  \\
  \item[(M.3)'']
\hskip.2cm{\it There exist constants $C,L>0$ such that}\\
 $ p^{\frac{p}{2}}\leq C\; L^p M_p,  \quad p = 0,1,\dots
$\\
({\it non triviality condition for the  spaces ${{\mathcal
S}^{\{M_p\}}}({\Bbb R}^d)$})
\end{itemize}

 When we discus spaces of Beurling type,
 instead of (M.3)'' we assume moreover:
 \begin{itemize}
\item[(M.3)''']
 {\it For every   $L>0$, there exists $C>0$ such that}\\
$ p^{\frac{p}{2}}\leq CL^pM_p,\quad  p = 1,2,\dots .
$\\
({\it non triviality condition for the spaces ${{\mathcal
S}^{(M_p)}}({\Bbb R}^d)$}).
\end{itemize}

We will denote by ${{\mathcal S}^{\{M_{p}\}}}'({\Bbb R}^d)$
(resp.\ ${{\mathcal S}^{(M_{p})}}'({\Bbb R}^d)$) the strong dual
of the space ${\mathcal S}^{\{M_{p}\}}({\Bbb R}^d)$ (resp.\
${{\mathcal S}^{(M_{p})}}({\Bbb R}^d)$). \\

 By
$$
\HH_n(x) =  \frac{(-1)^n }{\sqrt{2^{n}n!\sqrt{\pi}}} \:
e^{x^2/2}\frac{d^n}{dx^n}\left(e^{-x^2}\right), \quad n \in {\Bbb
N},
$$
we denote the {\it Hermite functions}. The functions are
eigenvalues of harmonic oscillator:
 $$\left(- \frac{d^2}{dx^2}+x^2 -1  \right)\HH_n = n\HH_n, $$
and that satisfy estimation
 estimation:
 $$ \| \HH_n \|_{\infty}\leq C n^k,   $$
for some $C$ and $k$ independent of $n$,
 They  are elements of the Gelfand-Shilov
spaces (see \cite{L} and \cite{nas rad}).\\

In multidimensional case the Hermite functions  are defined by
$$
\HH_n(x)= \HH_{n_1}(x_1)\HH_{n_2}(x_2)\cdots \HH_{n_d}(x_d), \quad
x =(x_1,x_2,...x_d)\in {\Bbb R}^d,
$$
where $n=(n_1,n_2,...,n_d) \in {\Bbb N}^d$. By $\HH_{(n,k)}$,
$(n,k)\in {\Bbb N}^l_0\times {\Bbb N}^s_0$, we denote
$$
\HH_{(n,k)}=\HH_{n_1}(x_1)\HH_{n_2}(x_2)\cdots \HH_{n_l}(x_l)
\HH_{k_1}(x_{l+1})\HH_{k_2}(x_{l+2})\cdots \HH_{k_s}(x_{l+s}).
$$

  The  Fourier-Hermite coefficients of an element $\varphi$ of Gelfand-Shilov space  are numbers
defined by:
$$a_n(\varphi) =\int_{{\Bbb R}^d} \varphi(x) \HH_n(x)dx, \quad
n \in {\Bbb N}^d_0,$$ and the  Fourier-Hermite coefficients of an
element $f$ of the dual space Gelfand-Shilov space  is defined by:
$$ b_n(f)= \langle f,\HH_n\rangle . $$
The sequence of the Fourier-Hermite coefficients
$\{a_n(\varphi)\}_{n\in {\Bbb N}^d_0}$ and $\{b_n(f)\}_{n\in {\Bbb
N}^d_0}$ of $\varphi$ and of $f$ are  the  Hermite representation
of $\varphi$ and $f$.

The proof of our theorems heavily rely on the characterization of
Fourier-Hermite coefficients of elements of Gelfand-Shilov spaces
and their duals.  Let us give a brief account on  the
characterization The test space ${{\mathcal S}^{\{M_p\}}}({\Bbb
R}^d) $ for Gelfand-Shilov space of Roumieu type (resp.\
${{\mathcal S}^{(M_p)}}({\Bbb R}^d) $ of Beurling) can be
identified with the space of multisequences of ultrafast falloff,
i.e. of multisequences $\{a_n\}_{n\in{\Bbb N}^d_0}$ of complex
numbers which satisfies that which satisfy that for some (resp.
each) $\theta =(\theta_1,..., \theta_n))$, where $ \theta_k > 0$,
for $k=1,...,d$,
$$ \left(\sum_{n \in {\Bbb N}_0^d} |a_n|^2 \exp \left[\sum_{k=1}^d M(\theta_k
\sqrt{n_k})\right]\right)^{1/2}< \infty,
$$

Here, $M(\cdot)$ is the associated function for the sequence
$\{M_p\}_{p\in {\Bbb N}_0}$ defined by
 \begin{equation}\label{asoc}
 M(\rho) =
        \sup_{p \in {\Bbb N}_0} \; \log
        \frac{\rho^{p}}{M_{p}}, \quad \rho > 0.
   \end{equation}
In the special case (\ref{Mp}), one have $ M(\rho)=
\rho^{\frac{1}{s}}(\log\,\rho)^{-\frac{t}{s}},  $  $\rho \gg 0$.\\

Since the space of sequences of ultrafast falloff is a Frechet
space, the spaces ${{\mathcal S}^{\{M_p\}}} $ and ${{\mathcal
S}^{(M_p)}}$ are Frechet spaces too.\\

The dual space ${{\mathcal S}^{\{M_p\}}}'({\Bbb R}^d) $  (resp.\
${{\mathcal S}^{(M_p)'}}({\Bbb R}^d)$) can be identified with the
space of multisequences of ultrafast growth. This implies that $f$
belongs to the space if and only if is Hermite representation
$\{b_n\}$ satisfies that
 for every (resp.\ some)
$\theta =(\theta_1,...\theta_d)\in{\Bbb R}^d_+$ holds:
\begin{equation}
|b_n(f)|\leq \exp\left[\sum_{k=1}^dM(\theta_k\sqrt{n_k})\right],
\quad n=(n_1,...n_d)
\end{equation}

One of the immediate consequences of the characterization  is the
 Parseval equation: For $f\in {{\mathcal S}^{*}}'$ and
$\varphi\in {\mathcal S}^{*}$ we have
  $$\langle f,\varphi\rangle =\sum_{n\in {\Bbb N}^d_0}b_n(f)a_n(\varphi),\quad
\varphi\in{\mathcal S}^{*}({\Bbb R}^d),
 $$
  where the multisequences
 $\{a_n(\varphi)\}_{n \in {\Bbb N}^d_0}$ and  $\{b_n(\varphi)\}_{n \in {\Bbb N}^d_0}$  are the Hermite
representatives of $\varphi $ and $f$.

Let us now prove the other important consequences of the
characterization.

\section{Main results}\label{main results}

\subsection{Regularity theorem}

\begin{theorem}
If $T$ belongs to the space $\mathcal{S}^{ \{ M_{p} \}'}$, then
for every $\mu > 0$ there exists a continuous and bounded function
$f$ such that in the space $\mathcal{S}^{ \{ M_{p} \}'}$:
\begin{equation} \label{eq:1}
    T = \sum_{\beta = 0}^{\infty}
    \frac{\mu^{2\beta}}{M_{2\beta}} \frac{1}{2^{\beta}} \left(- \frac{d^2}{dx^2}+ x^2 -1  \right)^{\beta}f.
\end{equation}
\end{theorem}

{\bf Proof.} Let $T$ be an element of the space $\mathcal{S}^{ \{
M_{p} \}'}(\mathbb{R})$ and let $\{ b_{n} \}_{n=0}^{\infty}$ be
its Hermite representation. It follows that for every $\theta >0$
we have:
\begin{equation} \label{eq:2}
    |b_{n}| \leq C \exp[M(\theta \sqrt{n})].
\end{equation}

Let $\mu$ be an arbitrary positive number and let $\{ a_{n}
\}_{n=0}^{\infty}$ be a sequence defined by:
\begin{displaymath}
    a_{n}=b_{n} \left(\sum_{\alpha=0}^{\infty}
    \frac{\mu^{2\alpha} n^{\alpha}}{M_{2\alpha}} \right)^{-1}.
\end{displaymath}

Note that the sequence $\{ a_{n} \}_{n=0}^{\infty}$ is well
defined sequence since by $(M.3)''$ we have:
\begin{eqnarray*}
    1 \leq \sum_{\alpha=0}^{\infty}
    \frac{\mu^{2\alpha} n^{\alpha}}{M_{2\alpha}} \leq C \sum_{\alpha=0}^{\infty}
    \frac{\mu^{2\alpha} n^{\alpha} L^{2\alpha}}{(2
    \alpha)^{\alpha}}
    \leq C  \sum_{\alpha=0}^{\infty} \frac{1}{\alpha^{2}} <
    \infty.
\end{eqnarray*}

First we will prove that
\begin{displaymath}
    \left|a_{n} \mathcal{H}_{n}\right| \leq C \frac{1}{n^{2}},
\end{displaymath}
which will imply (see for example \cite[ Th 7.10 and
Th.7.12]{Rudin}) that the sum $\sum_{n=0}^{\infty} a_{n}
\mathcal{H}_{n}$ uniformly converges to a bounded and continuous
function $f$.

Inequality (\ref{eq:2}), the estimate for Hermite functions and
the condition $(M.2)$ imply that for $\theta = \frac{\mu}{H}$ we
have:
\begin{eqnarray*}
    |a_{n} \mathcal{H}_{n}| &=& |b_{n}|\cdot \left(\sum_{\alpha=0}^{\infty}
    \frac{\mu^{2\alpha}}{M_{2\alpha}}n^{\alpha}\right)^{-1} \cdot
    |\mathcal{H}_{n}|\\
    & \leq & C \exp[M(\theta \sqrt{n})] \cdot \left(\sum_{\alpha=0}^{\infty}
    \frac{\mu^{2\alpha} n^{\alpha}}{H^{2 \alpha}
    M_{\alpha}^{2}} \right)^{-1} \cdot n^{k}\\
    & \leq & C \exp[M(\theta \sqrt{n})]\cdot \left(\sup_{\alpha} \frac{(\mu / H)^{2 \alpha}
    n^{\alpha}}{M_{\alpha}^{2}} \right)^{-1} \cdot  n^{k}\\
    & = & C \exp[M(\theta \sqrt{n}) - 2M(\mu/H \sqrt{n})] \cdot
    n^{k}\\
    &=& C \exp [- M(\mu/H \sqrt{n})] \cdot n^{k}
    = C \frac{n^{k}}{\sup_{\alpha} \frac{(\mu / H)^{\alpha}
    n^{\alpha/2}}{M_{\alpha}}}\\
    &\leq & C \frac{ n^{k} M_{2k+4}}{(\mu/H)^{2k+4} n^{k+2}}
    \leq C \frac{1}{n^{2}}.
\end{eqnarray*}

Next we show that in the space $\mathcal{S}^{ \{ M_{p}
\}'}(\mathbb{R})$ holds:
\begin{displaymath}
     T = \sum_{\beta = 0}^{\infty} \frac{\mu^{2\beta}}{M_{2\beta}}
     \frac{1}{2^{\beta}} \left(- \frac{d^2}{dx^2} + x^2 -1 \right)^{\beta}f.
\end{displaymath}

But first let us note that $(- \frac{d^2}{dx^2} + x^2 -1 )^{\beta}
f = \sum_{n=0}^{\infty} a_{n} \: n^{\beta} \: \mathcal{H}_{n}$.
This follows from the fact that $f$ is continuous and bounded
function (thus $f \in \mathcal{S}^{ \{ M_{p} \}'}(\mathbb{R})$, so
$(- \frac{d^2}{dx^2}  + x^2 -1 )^{\beta} f \in \mathcal{S}^{ \{
M_{p} \}'}(\mathbb{R})$), and from equality:
\begin{eqnarray}
    \Bigg\langle \left(- \frac{d^2}{dx^2} + x^2 -1\right) f , \mathcal{H}_{n}\Bigg\rangle &=&
    \langle f ,(- \frac{d^2}{dx^2} + x^2 -1 ) f \mathcal{H}_{n} \rangle  \\
    & = & \langle f ,  n \mathcal{H}_{n} \rangle = n \langle f ,\;  \mathcal{H}_{n}
    \rangle = n a_{n}. \nonumber
\end{eqnarray}

So, in the space $\mathcal{S}^{ \{ M_{p} \}'}(\mathbb{R})$ we
have:
\begin{eqnarray*}
    \sum_{\beta = 0}^{\infty} \frac{\mu^{2\beta}}{M_{2\beta}}
        \left(- \frac{d^2}{dx^2}  + x^2 -1\right)^\beta f
        &=& \sum_{\beta = 0}^{\infty}\frac{\mu^{2\beta}}{M_{2\beta}}
            \left(- \frac{d^2}{dx^2}  + x^2 -1\right)^{\beta}
            \sum_{n=0}^{\infty} a_{n} \mathcal{H}_{n}\\
        &=& \sum_{\beta = 0}^{\infty} \frac{\mu^{2\beta}}{M_{2\beta}}
            \sum_{n=0}^{\infty}a_{n} n^{\beta} \mathcal{H}_{n}\\
        &=& \sum_{\beta = 0}^{\infty} \sum_{n=0}^{\infty}
            \frac{\mu^{2\beta}}{M_{2\beta}}a_{n} n^{\beta} \mathcal{H}_{n}
\end{eqnarray*}

  and
\begin{eqnarray*}
    T &=& \sum_{n=0}^{\infty} b_{n} \mathcal{H}_{n} =
    \sum_{n=0}^{\infty} a_{n} \: \left(\sum_{\beta=0}^{\infty}
    \frac{\mu^{2\beta}}{M_{2\beta}}n^{\beta}\right) \:
    \mathcal{H}_{n} \\
    &=& \sum_{n=0}^{\infty} \sum_{\beta = 0}^{\infty}
    \frac{\mu^{2\beta}}{M_{2\beta}} \: a_{n} \: n^{\beta}
    \: \mathcal{H}_{n}.
\end{eqnarray*}

  It only remains to show that the above two sequences of
equations are equal. For every $\varphi \in \mathcal{S}^{ \{ M_{p}
\}}$ we have:
\begin{eqnarray*}
    \sum_{\beta = 0}^{\infty} \left|\frac{\mu^{2\beta}}{M_{2\beta}}
        \: a_{n} \: n^{\beta} \langle \mathcal{H}_{n},  \varphi \rangle \right|
        &=& |a_{n}| \: \left(\sum_{\beta = 0}^{\infty}
            \frac{\mu^{2\beta}}{M_{2\beta}} \: n^{\beta}\right)
            \: \left|\langle \mathcal{H}_{n}, \varphi  \rangle\right|\\
    &=& |b_{n}| \cdot |\langle \mathcal{H}_{n}, \hskip.1cm \varphi \rangle |
\end{eqnarray*}

  and therefore, since $\{b_{n}\}_{n=0}^{\infty}$ is the
Hermite representation of $T \in \mathcal{S}^{ \{ M_{p} \}'}$ and
${\langle \mathcal{H}_{n}, \varphi \rangle}_{n=0}^{\infty}$ is the
Hermite representation of $\varphi$, we have:
\begin{eqnarray*}
    \sum_{n=0}^{\infty} |b_{n}| \: |\langle \mathcal{H}_{n},  \varphi
    \rangle | \leq   \left(\sum_{n=0}^{\infty} |b_{n}|^{2} \:  \exp[-2M(\theta \sqrt{n})]\right)^{1/2}
     \cdot \\ \cdot \left(\sum_{n=0}^{\infty} |\langle \mathcal{H}_{n},  \varphi
    \rangle| \cdot \exp[2M(\theta \sqrt{n})]\right)^{1/2} < \infty.
\end{eqnarray*}

  So by \cite[Th.8.3.]{Rudin} we have that $(6)$ and $(8)$
are equal.
\begin{flushright}
 $\Box$
\end{flushright}

\subsection{Kernel Theorems}

In the section we will prove the kernel theorem for Gelfand-Shilov
spaces.  As a consequence of the Kernel theorem we
 have that the Weyl transform can be  extended  on a spaces of
 Gelfand-Shilov and this
 gives the possibility to introduce the $\Psi$DO theory
 in the framework of the spaces.

\begin{theorem}[{\bf Kernel theorem}] Every jointly continuous bilinear
 functional $K$ on
  $ {{\mathcal S}^{\{M_p\}}}({\Bbb R}^l)\times {{\mathcal
 S}^{\{M_p\}}}({\Bbb R}^s)$ defines a linear map ${\mathcal K} : {\mathcal
 S}^{\{M_p\}}({\Bbb R}^s)\rightarrow {{\mathcal
 S}^{\{M_p\}}}'({\Bbb R}^l)$ by
\begin{equation}\label{kraj}
\langle {\mathcal K}\varphi, \psi \rangle = K(\psi
\otimes\varphi),\quad \text{for}\;\; \psi\in {\mathcal
S}^{\{M_p\}}({\Bbb R}^l), \varphi \in {\mathcal S}^{\{M_p\}}({\Bbb
R}^s)
\end{equation}
and $\big(\varphi\otimes\psi\big)(x,y)=\varphi(x)\psi(y)$, which
is continuous in the sense that ${\mathcal K} \varphi_j
\rightarrow 0 $ in ${{\mathcal
 S}^{\{M_p\}}}'({\Bbb R}^l)$ if $ \varphi_j \rightarrow 0 $ in ${{\mathcal
 S}^{\{M_p\}}}({\Bbb R}^s)$.

 Conversely, for every such linear map ${\mathcal K}$ there is a unique
 $K \in {{\mathcal S}^{\{M_p\}}}'({\Bbb R}^{l+s})$ such that
(\ref{kraj}) is valid.
 The $K $ is called the kernel of operator ${\mathcal K}$.
 \end{theorem}

{\bf Proof.}  If $K$ is a jointly continuous bilinear functional $
\in {{\mathcal S}^{\{M_p\}}}({\Bbb R}^l)\times {{\mathcal
 S}^{\{M_p\}}}({\Bbb R}^s)$, then (\ref{kraj}) defines a an element
 $({\mathcal K}\varphi) \in {{\mathcal S}^{\{M_p\}}}'({\Bbb R}^l)$ since
 $
 \psi \mapsto K(\psi\otimes\varphi)
 $
is continuous. The mapping ${\mathcal K}:{{\mathcal
S}^{\{M_p\}}}({\Bbb R}^{s})\times {{\mathcal S}^{\{M_p\}}}'({\Bbb
R}^{l})$ is continuous since the mapping $ \varphi \mapsto
K(\psi\otimes\varphi)
 $
is continuous.\\

Let us prove the converse. To prove the existence we define a
bilinear form $B$ on ${{\mathcal S}^{\{M_p\}}}'({\Bbb
R}^{l})\otimes {{\mathcal S}^{\{M_p\}}}'({\Bbb R}^{s})$ by
$$
B(\varphi,\psi) = \langle{\mathcal K}\psi,\phi\rangle, \quad
\psi\in {\mathcal S}^{\{M_p\}}({\Bbb R}^l),\; \varphi \in
{\mathcal S}^{\{M_p\}}({\Bbb R}^s).
$$
The form $B$ is a separately continuous  bilinear form on the
product ${\mathcal S}^{\{M_p\}}({\Bbb R}^l)\times {\mathcal
S}^{\{M_p\}}({\Bbb R}^s)$ of Frechet spaces and therefore it is
jointly continuous, see \cite{Trev}.\\

 Let $C>0$, $\theta \in {\Bbb R}^l_+$, $\nu \in {\Bbb R}^s_+$ be chosen so that
\begin{equation}\label{B1}
|B(\varphi,\psi)|\leq C||\varphi||_{\theta}||\psi||_{\nu},
\end{equation}
and let
\begin{equation*}
t_{(n,k)}= B(\HH_n,\HH_k), \quad n \in {\Bbb N}^l,k\in {\Bbb N}^s.
\end{equation*}
Since $B$ is jointly continuous on ${{\mathcal S}^{\{M_p\}}}({\Bbb
R}^l)\times {{\mathcal
 S}^{\{M_p\}}}({\Bbb R}^s)$, for
 $\varphi=\sum a_n \HH_n$ and
$\psi=\sum b_k \HH_k$ we have that
\begin{equation*}
B(\varphi,\psi)=\sum t_{(n,k)}a_n b_k.
\end{equation*}
On the other hand, for $ (n,k) \in {\Bbb N}^l \times {\Bbb N}^s$
and $(\theta,\nu)\in {\Bbb R}^l \times{\Bbb R}^s$, by (\ref{B1})
we have
\begin{equation*}
\begin{split}
    |t_{(n,k)}| & \leq
        C||\HH_{n}||_{\theta}||\HH_k||_{\nu}=\\
        &= \left||\HH_{n_1}\right||_{\theta_1}
            ||\HH_{n_2}||_{\theta_2}\cdots ||\HH_{n_l}||_{\theta_l}\;
            ||\HH_{k_1}||_{\nu_1} ||\HH_{k_2}||_{\nu_2}
            \cdots||\HH_{k_s}||_{\nu_s}=\\
        &=\exp\left[2\sum_{i=1}^{l}M\big(\theta_i\sqrt{n_i}\big)\right]
            \exp\left[2\sum_{j=1}^{s}M\big(\nu_j\sqrt{k_j}\big)\right].
\end{split}
\end{equation*}

Thus, from Hermite characterizations of Gelfand-Shilov spaces, it
follows that the sequence $\{t_{(n,k)}\}_{(n,k)}$ is a Hermite
representation of a an element $K\in{{\mathcal
S}^{\{M_p\}}}'({\Bbb R}^l\times {\Bbb R}^s)$. Thus
\begin{equation}\label{dosta}
\langle K,\varphi\rangle = \sum t_{(n,k)}c_{(n,k)},
\end{equation}
for  $\varphi=\sum c_{(n,k)}\HH_{n,k}\in {\mathcal
S}^{\{M_p\}}({\Bbb R}^{l+s})$.\\

If $\varphi=\sum a_n \HH_n \in {\mathcal S}^{\{M_p\}}({\Bbb
R}^{l})$ and $\psi=\sum b_k \HH_k \in {\mathcal S}^{\{M_p\}}({\Bbb
R}^{s})$ then $\varphi\otimes\psi$ has the Hermite representation
$\{a_nb_k\}_{(n,k)}$ and we have that for $K$ defined by
(\ref{dosta})
$$
K(\varphi\otimes\psi)=\sum_{(n,k)}t_{(n,k)}a_nb_k =
B(\varphi,\psi),
$$
so $K=B$.  This proves the existence.

The uniqueness follows from the fact that $K$ is completely
determined by its Hermite representation $\{\langle
K,\HH_{(n,k)}\rangle \}_{(n,k)}$ and the fact that for every
$(n,k)\in {\Bbb N}^l \times {\Bbb N}^s $
$$
\langle K,\HH_{(n,k)}\rangle = \langle K,\HH_n\otimes \HH_k\rangle
= B(\HH_n,\HH_k) = t_{(n,k)}.
$$
\begin{flushright}
 $\Box$
\end{flushright}

\section{Remarks}\label{remarks}

\subsection{Examples of Gelfand-Shilov spaces}
The Gelfand Shilov spaces generated by the sequences $\{M_p\}$,
which belong to the classes of sequences  considered in the paper,
are generalizations of many spaces known in the literature.

\begin{itemize}
\item
If $M_p=p^{\alpha p}$,  the space ${{\mathcal
S}^{\{M_{p}\}}}({\Bbb R}^d)$ is the Gelfand-Shilov space
${{\mathcal S}^{\alpha}_{\alpha}}$ (\cite{Gelfand2})
 and ${{\mathcal S}^{(M_{p})}}({\Bbb R}^d)$ is the Pilipovi\'c
space ${\sum_{\alpha}^{\alpha}}$ (\cite{PT}).
\item
If $M_p=p^p$ then the space ${{\mathcal S}^{\{M_{p}\}}}({\Bbb
R}^d)$ is isomorphic with the Sato space ${\mathcal F}$, the  test
space for Fourier hyperfunctions ${\mathcal F}'$, and ${{\mathcal
S}^{(M_{p})}}({\Bbb R}^d)$ is the Silva space ${\mathcal G}$, the
test space for extended Fourier hyperfunctions ${\mathcal G}'$ .
\item
The space considered in \cite{Fizika} and \cite{Fizika2} is
isomorphic with ${\mathcal S}^{\{p!\}}$.
\item
Braun-Meise-Taylor space ${\mathcal S}_{\{\omega\}}$, $\omega \in
{\mathcal W}$,   studied in the series of papers by the same
authors (see \cite{Braun2}and references in there), is the space
${\mathcal S}^{\{M_p\}}({\Bbb R}^d)$, where
$$M_p=\sup_{\rho>0}\rho^{p}e^{-\omega(\rho)}.$$ The sequence
satisfies the conditions (M.1), (M.2) and (M.3)', and it is in
general different  from a Gevrey sequence.
\item
Beurling-Bj\"ork space ${\mathcal S}_{\omega}$, $\omega \in
{\mathcal M}_c$, introduced in \cite{BJ}, is equal to the space
${\mathcal S}^{(M_p)}({\Bbb R}^d)$, where  $$M_p=
\sup_{\rho>0}\rho^pe^{-\omega(\rho)}.$$  The sequence satisfies
the conditions (M.1) and (M.3)', and it is in general different
from a Gevrey sequence. If we assume additionally that
$\omega(\rho) \geq C(\log \rho)^2$ for some $C>0$, then (M.2) is
satisfied.

\item  In \cite{Kor} Korevaar developed a very general theory of
Fourier transforms,  based on a set of original and well motivated
ideas. In order to obtain a formal class of objects which contain
functions of exponential growth and which is closed under Fourier
transform he introduced objects called pansions of exponential
growth. From characterization theorem \cite[Theorem 92.1]{Kor} and
\cite{L} it follows that exponential pansions are exactly Gelfand
Shilov spaces of Roumieu type, generated by the sequence
$M_p=p^{p/2}$.
\end{itemize}

\subsection{About conditions we assume on $\{M_p\}$}
Let us give few remarks about the conditions we assume on the
sequence $\{M_p\}$.

 The condition (M.1) is of technical nature,
which simplify the work and involve no loss of generality. This is
the well known fact for the Denjoy-Carleman classes of functions.
See for example \cite{Rudin2}.

The  condition (M.2) is standard in the  theory of generalized
functions (see \cite{K}). It implies that the class
${C^{\{M_p\}}}$ is closed under the (ultra)diffe\-ren\-tia\-tion,
 and is important in characterization of Denjoy-Carleman classes in
multidimensional case.

To be able to discuss our results in the context of Komatsu's
ultradistributions, let us state condition (which we do not
assume) :
\begin{itemize}
 \item[(M.3)'\hskip-.2cm]
\hskip.2cm$ \sum^{\infty }_{p=1}
     \frac{M_{p-1}}{M_{p}} < \infty.$
 \\({\it non-quasi-analyticity})
\end{itemize}

The non-triviality  conditions (M.3)'' and (M.3)'''  are weaker
then the condition (M.3)'. Under the conditions (M.3)"  and
(M.3)'''. For example, the sequence (\ref{Mp}) satisfies
conditions (M.1), (M.2), (M.3)", and if $t>0$ also the condition
(M.3)''' but not (M.3)'; and for $s > 1$ it satisfies the stronger
condition
(M.3)'.\\

The condition (M.3)' is necessary and sufficient condition that
the classe ${C^{\{M_p\}}} $ has a nontrivial subclass of functions
with compact support, i.e. that ${C^{\{M_p\}}} $ is
non-quasianalytic class of functions. This is standard
nontrivially condition in the ultradistribution theory. If it is
satisfied Gelfand-Shilov spaces are  the proper subspaces of the
Komatsu's spaces of ultradistributions (\cite{K}) of Roumieu or
Beurling type, so-called tempered ultrdistributions. If it is not
satisfied the Komatsu's spaces are trivial and elements of
Gelfand-Shilov spaces are
hyperfunctions.\\

The smallest nontrivial Gelfand-Shilov space is ${\mathcal
S}^{1/2}_{1/2}$. Condition (M.3)" essentially means that the space
${\mathcal S}^{1/2}_{1/2} $ is a subset of ${{\mathcal
S}^{\{M_p\}}} $.\\

The smallest nontrivial Pilipovi\'c space
${\sum_{\alpha}^{\alpha}}$ (\cite{PT}) does not exist. Note,
$\sum_{1/2}^{1/2}=\{0\}$, but the space $\sum_{\alpha}^{\alpha} $,
$\alpha > 1/2$, is nontrivial. Moreover, every nontrivial
Pilipovi\'c space $\sum_{\alpha}^{\alpha} $, contains as a proper
subspace one Gelfand Shilov space of Beurling type, for example,
the space ${{\mathcal S}^{(M_p)}} $, where $M_p= p^{p/2}(\log
p)^{pt}$, $t>0$.

\bibliographystyle{amsplain}

\end{document}